\documentclass{article}

\usepackage{microtype}
\usepackage{graphicx}
\usepackage{url}
\usepackage{subfigure}
\usepackage{booktabs}
\usepackage{array}
\usepackage{amsmath}
\usepackage{amssymb}

\newcommand{\msinfer}{\textsc{MSInfer}}
\newcommand{\gdn}{Gated DeltaNet}
\newcommand{\agentassist}{Agent-Assisted}

\usepackage[accepted]{mlsys2026}
\usepackage[hidelinks]{hyperref}

\makeatletter
\renewcommand{\Notice@String}{Technical report.}
\makeatother

\mlsystitlerunning{AI-Assisted Gated DeltaNet Optimization on NVIDIA Blackwell}

\begin{document}

\twocolumn[
\mlsystitle{Technical Report: AI-Assisted Gated DeltaNet Optimization on NVIDIA Blackwell}

\mlsyssetsymbol{equal}{*}

\begin{mlsysauthorlist}
\mlsysauthor{Hyunjun Shin}{ind}
\mlsysauthor{Jiseung Jang}{ind}
\mlsysauthor{Jaewoo Maeng}{ind}
\mlsysauthor{Hyunjun Kim}{ind}
\end{mlsysauthorlist}

\mlsysaffiliation{ind}{Independent Researcher}

\mlsyscorrespondingauthor{Hyunjun Shin}{shjj1504@gmail.com}
\mlsyscorrespondingauthor{Jiseung Jang}{jsg1504@gmail.com}
\mlsyscorrespondingauthor{Jaewoo Maeng}{jwmaeng@snu.ac.kr}
\mlsyscorrespondingauthor{Hyunjun Kim}{hjun20.kim@gmail.com}

\mlsyskeywords{GPU systems, kernel optimization, AI-assisted programming, Blackwell, FlashInfer, Gated DeltaNet}

\vskip 0.3in

\begin{abstract}
AI-assisted GPU programming is often framed as a kernel-generation loop: ask a model to produce faster CUDA code, benchmark the result, and repeat. This case study argues that contest-grade optimization involves more than improving the kernel body. We examine the \agentassist{} submission by our team, \msinfer{}, to the MLSys 2026 FlashInfer Contest. The submission optimized \gdn{} decode and prefill on NVIDIA B200/Blackwell and achieved an official 1.58$\times$ speedup, with approximate average latencies of 9.315~$\mu$s for decode and 239.48~$\mu$s for prefill. Our experience shows that even effective local kernel improvements can plateau when a workload requires structural reformulation and evaluator-aligned measurement. We therefore characterize AI-assisted kernel optimization as an end-to-end systems problem that encompasses algorithm design, workload specialization, measurement tooling, build and evaluation surfaces, evaluator alignment, and human interpretation.
\end{abstract}
]

\printAffiliationsAndNotice{}

\section{Introduction}

AI-assisted systems programming promises to shorten the path from an optimization idea to executable GPU code. Large language models can propose CUDA kernels, rewrite indexing logic, generate benchmark scripts, and summarize profiling output. In a contest setting, however, the useful unit of optimization is not the kernel body alone. The submitted artifact must satisfy the evaluator's build contract, run on the target architecture, pass all correctness workloads, and measure well under the same baseline and timing conditions used for ranking.

This paper studies that broader system through the \agentassist{} submission by our team, \msinfer{}, to the MLSys 2026 FlashInfer Contest~\cite{flashinfer2026contest}. The submission targeted \gdn{} (GDN) decode and prefill on NVIDIA B200/Blackwell. Our team produced a correct final submission with an official 1.58$\times$ \gdn{} speedup. Its final approximate latencies were 9.315~$\mu$s for decode and 239.48~$\mu$s for prefill, with all 54 decode and 100 prefill evaluation workloads passing correctness.

This report is not a winner's retrospective. The central observation is instead that our team made meaningful progress through disciplined local search, but the prefill path remained close to a sequential recurrence. That outcome is useful because it exposes where AI-assisted micro-optimization stops being enough: the search loop can improve arithmetic, data movement, launch settings, and measurement discipline, yet still miss a larger change in computational shape.

Measurement and algorithm design were coupled throughout the project. A workload-specialized prefill strategy only pays off if the team can reliably tell which workload regimes improved and which regressed. The inspected \msinfer{} workflow accumulated Modal benchmark histories, NCU profiles, scorecards, workload inventories, and phase timers, but it did not center a single official-adjacent kernel-latency measurement primitive during most of the search. The result was a broad evidence base that helped reject many local ideas but provided a weaker signal for promoting workload-specific structural rewrites.

The central thesis is:

\begin{quote}
Disciplined measurement and AI-assisted micro-optimization are necessary for contest-grade GPU work, but for \gdn{} prefill they were not sufficient. The \msinfer{} case suggests that structural reformulation, workload-shape specialization, and evaluator-aligned measurement must be treated as first-class parts of the optimization loop.
\end{quote}

This case study makes five contributions.
\begin{enumerate}
\item It records the final \agentassist{} \gdn{} submission by \msinfer{}~\cite{bammuri2026agentdecode,bammuri2026agentprefill,bammuri2026msinfer}, including official speedup, approximate decode and prefill latencies, correctness coverage, and the artifact boundary used for interpretation.
\item It analyzes the team's kernel and runtime choices, including algebraic simplification, V-dimension splitting, destination passing, \texttt{cp.async}, NCU-driven reasoning, and B200-oriented runtime pinning.
\item It extracts lessons from public high-performing \agentassist{} artifacts, separating decode as a shape-dispatch and launch/runtime specialization problem from prefill as a structural chunk/WY/tensor-core reformulation problem.
\item It analyzes evaluator alignment and measurement tooling as optimization signals, including a Modal setting where \msinfer{} decode appeared faster than FlashInfer, while an official B200 extra-round evaluation measured a much stronger baseline and reversed the conclusion.
\item It provides an evidence audit that separates official \agentassist{} artifacts, public comparison artifacts, internal measurement logs, and companion Full-Agent-style workflow artifacts, so that performance claims remain tied to the final submitted Agent-Assisted artifacts.
\end{enumerate}

Unless stated otherwise, public comparison values are drawn from the cited reports and were not independently reproduced; internal Modal measurements are reported separately from official results.

\section{Background}

\subsection{Contest Overview}

The MLSys 2026 FlashInfer Contest was an AI kernel-generation competition centered on submitting optimized GPU kernels for predefined FlashInfer benchmark definitions~\cite{flashinfer2026contest,flashinfer2026starterkit}. Participants submitted tagged Git repositories with per-definition configuration files, and the evaluator built and ran those submissions on the target platform. The contest distinguished a Full-Agent track, where the submitted agent must reproduce the kernel end to end, from an \agentassist{} track, where human experts and agents may collaborate and the team submits the resulting kernel code. This paper concerns only the \agentassist{} GDN submission.

The GDN track had two separate definitions: decode and prefill. The contest FAQ states that the final GDN score is the average speedup across the decode and prefill definitions, and that per-definition speedup is measured against a simple reference implementation rather than an optimized FlashInfer baseline. This distinction matters for the rest of the paper. A submission can report a favorable contest speedup while still being slower than a highly optimized FlashInfer baseline under a different evaluation setting, so we treat official score, absolute latency, correctness coverage, and benchmark source as separate pieces of evidence.

\subsection{GDN Decode and Prefill}

\gdn{} exposes two different optimization regimes. Decode is latency-sensitive and often operates at microsecond scale. Small changes in launch overhead, synchronization, compiler target, and baseline implementation can change the apparent outcome. Prefill processes longer sequences and larger token sets. It therefore offers more opportunity for chunking, tensor-core use, and shape-specialized scheduling, but it also requires handling a recurrent state update.

The \msinfer{} project worked with evolving evaluation workloads. The decode workload set grew to 54 workloads, while prefill used 100 workloads. Because the benchmark suite, wrapper versions, and baseline conditions changed during the project, early internal results were not directly comparable to final official results.

\subsection{Evaluation Surface}

The target platform was NVIDIA B200/Blackwell. We use \emph{build and evaluation surfaces} to refer to the compiler, runtime, binding, packaging, and measurement interfaces between a kernel and the evaluator. In this setting, \texttt{sm\_100a} targeting, language choice, runtime compilation, TVM FFI, destination-passing style, and dependency pins can change both the generated code and the evaluator's ability to load it. Our team therefore treated these interfaces as optimization concerns, not only as packaging details.

\section{MSInfer Submission}

\subsection{Official Result}

Table~\ref{tab:official-results} summarizes the final \msinfer{} \agentassist{} \gdn{} result. The official headline is a 1.58$\times$ speedup.

\begin{table}[t]
\caption{Final \msinfer{} \agentassist{} \gdn{} result. Latencies are approximate per-workload averages.}
\label{tab:official-results}
\vskip 0.05in
\begin{center}
\begin{small}
\setlength{\tabcolsep}{3pt}
\begin{tabular}{>{\raggedright\arraybackslash}p{1.1in}>{\raggedright\arraybackslash}p{1.45in}}
\toprule
Metric & Value \\
\midrule
Official \gdn{} speedup & 1.58$\times$ \\
Decode latency & 9.315~$\mu$s \\
Prefill latency & 239.48~$\mu$s \\
Decode correctness & 54/54 workloads \\
Prefill correctness & 100/100 workloads \\
Decode artifact & \texttt{69f517b} \\
Prefill artifact & \texttt{864f286} \\
\bottomrule
\end{tabular}
\end{small}
\end{center}
\vskip -0.1in
\end{table}

\subsection{Kernel Techniques}

\msinfer{} applied several useful optimizations. The shared algebraic simplification was to pre-scale state by the gate and compute output as
\[
  \mathrm{out} = q^\top(gS) + \delta(q^\top k),
\]
instead of recomputing a full dot product against the updated state. This made the decode and prefill paths share a useful form:
\[
\begin{aligned}
g &= \exp(-\exp(A_{\log}) \cdot \mathrm{softplus}(a + b_{\Delta t})), \\
\beta &= \sigma(b), \\
\mathrm{old}_v &= k^\top(gS), \\
\delta &= \beta(v - \mathrm{old}_v), \\
S' &= gS + k\delta^\top.
\end{aligned}
\]
For decode, this reduced redundant state reads and unnecessary scalar synchronization. For prefill, the same form removed a second full output dot per token in the sequential recurrence.

\msinfer{} also used V-dimension splitting, warp-local reductions, \texttt{cp.async.cg} style movement for prefill K/Q staging, destination-passing style, and runtime pinning around B200 compilation. These changes improved performance in internal measurements and produced an auditable record of the optimization process. However, they did not eliminate the remaining performance gap, especially in prefill.

\msinfer{} did make algebraic changes; it was not a purely mechanical implementation of the baseline recurrence. However, these changes were local reformulations inside a token-by-token execution model. The public-artifact comparison in Section~\ref{sec:comparative-evidence} points to a different level of computation: grouping multiple prefill tokens into chunks, summarizing the recurrence over each chunk, and exposing larger matrix-like operations that can use tensor cores. In short, \msinfer{} optimized the sequential recurrence path, while the missing direction changed the computational shape of prefill.

The decode basis for this report is the final \agentassist{} decode artifact at commit \texttt{69f517b}, tagged \texttt{gdn-decode}~\cite{bammuri2026agentdecode}. We keep the artifact identity explicit because non-final and Full-Agent branches also existed, but the main text focuses on the performance interpretation rather than listing every configuration field.

For the \agentassist{} decode artifact, the team followed a disciplined local-search process rather than adopting a late Full-Agent hybrid branch. The workflow imposed phase targets, one-change-at-a-time edits, immediate rollback on correctness or latency regression, and median-of-five promotion once Modal noise became visible. The inspected NCU log for the \agentassist{} decode path recorded a 54/54 Modal pass and an internal arithmetic mean around 12~$\mu$s before later finalization. An NCU capture at $B=64$ showed low issue-slot utilization, low achieved occupancy, and too few waves per SM; the bottleneck was insufficient parallelism and bytes in flight, not a single saturated HBM pipeline. This internal measurement is distinct from the official decode latency of 9.315~$\mu$s reported in this paper.

The prefill basis is the final \agentassist{} artifact at commit \texttt{864f286}, tagged \texttt{gdn-prefill}~\cite{bammuri2026agentprefill}. Its evaluator-facing configuration remains relevant to reproducibility, but the main comparison concerns the type of recurrence computation that the artifact did and did not implement.

For prefill, the plateau is clearer. The inspected internal and companion logs show large early gains from register-tiled state and loop fusion, from 34{,}280~$\mu$s to 11{,}510~$\mu$s, then V-split blocks from 8{,}260~$\mu$s to 5{,}210~$\mu$s, and warp-parallel V-row algebraic fusion from 5{,}210~$\mu$s to 2{,}110~$\mu$s. Later entries cluster around smaller or reverted changes: software prefetching, q/k reduction decoupling, fused \texttt{qk\_dot}, wider blocks, and additional split-factor tiers did not produce robust latency gains. This supports the main interpretation of the case study: local recurrence optimization produced real progress, then plateaued before a larger algorithmic reformulation.

\subsection{What MSInfer Did Not Do}

The most important limitation is that \msinfer{} mostly optimized a register-resident scalar recurrence. It improved arithmetic, data movement, synchronization, and work partitioning inside that recurrence, but it did not turn prefill into a chunk-level tensor-core problem. The later public comparison shows that high-performing artifacts did. This distinction is the main reason to frame the report as a differential case study rather than a simple contest narrative. It also avoids an overclaim: \msinfer{}'s algebraic simplification was meaningful, but it was a local per-token reformulation rather than the structural chunk/WY reformulation that changed the prefill workload into larger block computations.

\section{Measurement and Evaluator Alignment}

\subsection{Why Modal Speedup Was Misleading}

Speedup was not stable enough to be the only engineering metric. During development, a Modal run suggested that \msinfer{} decode beat FlashInfer: the baseline was approximately 29.3~$\mu$s and \msinfer{} was approximately 14.8~$\mu$s, or about 1.98$\times$. An official B200 extra-round evaluation told a different story: the FlashInfer baseline was approximately 6.55~$\mu$s and \msinfer{} was approximately 8.12~$\mu$s, or about 0.80$\times$ versus baseline.

\begin{table}[t]
\caption{Decode measurement divergence. Values are approximate and should be read as an evaluator-alignment lesson, not as a controlled causal experiment.}
\label{tab:modal-official}
\vskip 0.05in
\begin{center}
\begin{small}
\begin{tabular}{lccc}
\toprule
Setting & Baseline & \msinfer{} & Ratio \\
\midrule
Modal & 29.3~$\mu$s & 14.8~$\mu$s & 1.98$\times$ \\
Official B200 extra-round & 6.55~$\mu$s & 8.12~$\mu$s & 0.80$\times$ \\
\bottomrule
\end{tabular}
\end{small}
\end{center}
\vskip -0.1in
\end{table}

These numbers should not be read as a simple \msinfer{} regression. They show that the reference baseline, evaluator configuration, build path, clock conditions, and hardware environment can move the denominator dramatically. The safer engineering practice is to report absolute latency, pass counts, workload coverage, build configuration, and benchmark source for every number.

\subsection{Evaluation Tooling as Competitive Advantage}

The measurement gap did more than complicate post hoc analysis; it changed which optimizations looked promising. \msinfer{} improved its methodology with repeated runs, isolated runners, full-workload validation, and NCU profiles, but it still relied substantially on contest-facing or Modal-facing timing paths during search. Kachua's public repository exposes a more sharply separated timing stack: \texttt{run\_modal.py} for packed benchmark/correctness runs, \texttt{bench\_fi\_timing.py} and \texttt{bench\_fi\_timing\_modal.py} for FlashInfer/CUPTI-style pure GPU timing, and \texttt{profile\_modal.py} for profile artifacts. Its experiment instructions explicitly log both \texttt{benchmark\_latency} and \texttt{kernel\_latency} in microseconds and prefer the latter. This suggests that evaluation tooling itself contributed to contest performance.

This distinction matters because an optimizer follows the metric it can observe. If the local harness is noisy or misaligned, a human-agent loop can select variants that win the local measurement but fail to transfer. If the harness is stable and official-aligned, smaller workload-specific choices become actionable. The \msinfer{} tooling was broad but more fragmented: companion workflow tooling logged Modal B200 latency, speedup, correctness, and workload counts, while later agent repositories added precise/decision-gate benchmark modes, NCU scorecards, surrogate scorecards, workload inventories, and phase timers. These tools helped reject bad ideas and diagnose bottlenecks, but they did not give the search loop one central FlashInfer/CUPTI-style kernel-latency primitive comparable to Kachua's public timing path.

This is the most important tooling distinction in the case study. \msinfer{} did not lack engineering instrumentation; it had many forms of instrumentation. What it lacked was a single, trusted decision metric that made per-workload algorithm dispatch easy to rank. Kachua's repository suggests a different approach: correctness and benchmark scripts still existed, but the experimental loop could privilege raw kernel timing when deciding whether a candidate formulation should advance.

\subsection{Profiler Evidence}

NCU profiling pushed the team away from single-number reasoning. For decode, one B200 profile showed low issue-slot utilization and insufficient parallelism/bytes in flight rather than a single saturated HBM bottleneck. That matters because naive state-traffic reasoning would have labeled the kernel memory-bound too quickly. The profiler suggested a subtler small-kernel problem: too few waves and too little concurrent work for B200, especially when launch overhead and cache behavior were folded into the measurement.

For prefill, profiling and comparative evidence point to a different conclusion. The issue was not only that \msinfer{} needed another cache hint or fewer barriers. The missing piece was a formulation that exposes larger tensor-core-friendly chunks while preserving correctness.

\section{Comparative Evidence and Lessons}
\label{sec:comparative-evidence}

We now compare \msinfer{} with public high-performing \agentassist{} artifacts to extract design lessons rather than reconstruct the leaderboard. Table~\ref{tab:top-comparison} summarizes the public top-three comparison and separates reported numbers from design lessons. We tie all exact values to pinned public reports or repositories.

\begin{table*}[t]
\caption{Public top-three \agentassist{} GDN comparison. Reported values are drawn from the cited public artifacts.}
\label{tab:top-comparison}
\vskip 0.05in
\begin{center}
\begin{scriptsize}
\setlength{\tabcolsep}{3pt}
\begin{tabular}{>{\raggedright\arraybackslash}p{0.9in}>{\raggedright\arraybackslash}p{0.6in}>{\raggedright\arraybackslash}p{0.85in}>{\raggedright\arraybackslash}p{0.9in}>{\raggedright\arraybackslash}p{2.6in}}
\toprule
Submission & Rank & Decode & Prefill & Main design lesson \\
\midrule
Kachua & 1st & public repo raw trajectory: 5.44--5.47~$\mu$s median-of-workload medians for selected CUDA variants & public repo raw rows: 61.17--61.50~$\mu$s & Decode: direct TVM FFI, one-warp \texttt{BV=8} CUDA, vectorized loads, and butterfly reductions. Prefill: chunkwise matmul plus Neumann/WY-style inverse, adaptive \texttt{CHUNK}/\texttt{BV}, tensor-core \texttt{tl.dot}, and GPU-timing-aligned evaluation. \\
UW SyFI & 2nd & Not reported in the inspected public artifacts & Not reported in the inspected public artifacts & Decode: NVRTC/Driver API launch interface, TVM FFI, and one-warp \texttt{V\_TILE=8}. Prefill: BF16 WY-form chunked CUDA with WMMA/\texttt{cp.async}. The runtime interface was part of the optimization. \\
LLM-CUDA & 3rd & reported 6.201~$\mu$s & reported 51.992~$\mu$s & Decode: hybrid dispatch across small recurrent, B16 high-CTA CuTe, B48 \texttt{TILE\_V=16}, and default large-batch CuTe paths. Prefill: Blackwell chunk kernel plus narrow short-shape fallback. \\
\msinfer{} & outside top three & 9.315~$\mu$s & 239.48~$\mu$s & Passed all correctness workloads. Decode stayed mostly within one templated CUDA family selected by split factor; prefill stayed in a register-resident scalar recurrence family with V-split, warp rows, gate pipeline, and local micro-optimization. \\
\bottomrule
\end{tabular}
\end{scriptsize}
\end{center}
\vskip -0.1in
\end{table*}

\subsection{Kachua: Prefill as Algorithmic Reformulation}

Kachua is the most important comparison point for \msinfer{}~\cite{kachua2026agentassist,kachua2026repo}. Its public repository exposes multiple prefill families: Direct v2, Split-WY, and Flat-WY. The top-level kernel comment describes prefill as chunkwise matmul plus a Neumann inverse: each sequence is split into fixed-size chunks, the intra-chunk recurrence is summarized algebraically, and Tensor Core \texttt{tl.dot} is used for TF32/BF16/FP16 contractions. The key dispatch axes are \texttt{CHUNK}, the number of tokens grouped into a local recurrence solve, and \texttt{BV}, the number of state or output rows updated by one program. The code chooses these values from \texttt{num\_seqs}, total tokens, and an average-sequence-length proxy, then routes to Split-WY, Flat-WY, or Direct v2. This is qualitatively different from only changing block size or vector load width. The repository also gives raw microsecond evidence: prefill \texttt{results.csv} records accepted rows at 61.50, 61.20, and 61.17~$\mu$s, while the CUDA decode README reports selected variants with 5.44--5.47~$\mu$s median-of-workload medians. We treat these as public-repository trajectory numbers, not as independently reproduced official leaderboard averages.

This comparison shows that \gdn{} prefill can be improved by changing the mathematical and scheduling shape of the recurrence. Kachua converted more work into reusable chunk metadata and tensor-core-friendly local solves. \msinfer{} instead made a scalar recurrence faster. That distinction explains why \msinfer{} could improve internally yet remain far from top prefill latency.

Kachua's repository also shows that evaluation tooling was part of the design rather than an afterthought. The repo includes \texttt{bench\_fi\_timing.py}, whose header describes a FlashInfer pure-GPU timer that mirrors a CUPTI path and excludes Python dispatch and benchmark-framework overhead. Its Modal wrapper, \texttt{bench\_fi\_timing\_modal.py}, states that it matches the reference repository's Modal CUPTI path rather than the latency path used by \texttt{scripts/run\_modal.py}, installs \texttt{cupti-python}, and can sweep all workloads. The same repo's program notes say that \texttt{kernel\_latency} is more important than \texttt{benchmark\_latency} or speedup because it is the actual kernel latency. Thus the public repo supports a systems interpretation: Kachua appears to have optimized both the prefill formulation and the timing signal used to compare variants. The two reinforce each other. Structural prefill variants can be workload-sensitive, so a kernel-latency-centered harness makes it easier to decide where a chunked, WY-style, or fallback path is actually worth keeping.

\subsection{UW SyFI: Runtime Surface as a First-Class Variable}

UW SyFI's public artifacts emphasize another systems axis: the runtime and ABI surface~\cite{syfi2026agentassist,syfi2026repo}. Their decode path uses NVRTC and the CUDA Driver API to avoid a runtime/driver mismatch, while their GDN configurations use TVM FFI and destination-passing style. Their prefill CUDA source is explicitly a BF16 tensor-core chunked kernel in WY form, with \texttt{V\_TILE=16}, \texttt{C=16}, four-warp block specialization, WMMA fragments, and \texttt{cp.async.cg} double-buffered Q/K/V loads. The inspected public artifacts do not provide a pinned table of raw decode or prefill latencies; therefore, we do not report those values.

For \msinfer{}, the takeaway is not simply ``write CUDA instead of Python'' or ``use WMMA.'' The deeper lesson is that the evaluator-facing surface---language, binding, driver/runtime behavior, compiler target, and output allocation convention---is part of the optimization problem. This supports \msinfer{}'s own runtime-pin narrative around B200 and \texttt{sm\_100a}, but SyFI shows that such control must be paired with a strong prefill formulation.

\subsection{LLM-CUDA: Quantifying the Gap}

LLM-CUDA provides the clearest public numeric comparison~\cite{llmcuda2026agentassist,llmcuda2026repo}. Its retained \agentassist{} artifacts report about 6.201~$\mu$s for decode and 51.992~$\mu$s for prefill. Its prefill summary states that the retained candidate replaced an older wide pair-dispatch scheme with an upstream PR~\#3001 Blackwell chunk kernel as the default path, kept only a narrow recovered short-shape fallback for measured tiny-shape regressions, and passed a 100-workload official-aligned sweep. Relative to \msinfer{}'s official numbers, this makes \msinfer{} about 1.50$\times$ slower on decode and about 4.61$\times$ slower on prefill. The ratio is the main point. Decode was not solved, but prefill dominated the gap.

This comparison helps avoid a misleading conclusion. If we only inspect \msinfer{}'s decode story, the paper can sound like a near miss caused by evaluator noise. The prefill ratio says otherwise: the top-three path required a larger algorithmic move.

\subsection{Decode Gap: Dispatch and Shape Split}

The decode gap is better explained as a dispatch and small-kernel specialization gap than as a missing recurrence identity. \msinfer{}'s accepted \agentassist{} decode source is a single evaluator-facing CUDA/Python entry that compiles \texttt{kernel.cu} through \texttt{tvm\_ffi.cpp.load}. Inside the CUDA wrapper, \texttt{batch\_size} selects a \texttt{split\_factor}: small and mid batches use \texttt{split\_factor=8}, while larger batches use \texttt{split\_factor=4}; this maps to \texttt{ROWS\_PER\_WARP} instantiations of the same templated kernel. The kernel itself uses 128-thread blocks, four warps, vectorized state and q/k loads, shared staging for \texttt{v}, warp reductions, and L2 persistence hints~\cite{bammuri2026agentdecode}. This was a real decode implementation, but its shape split stayed within one kernel family.

The public top-three decode artifacts show a more aggressive separation of regimes. Kachua's CUDA decode README describes the retained direction as a direct TVM FFI C binding, a one-warp \texttt{BV=8} kernel, 128 CTAs for \texttt{B=1}, vectorized state and q/k/v loads, paired butterfly warp reductions, fused decay and delta-rule logic, interleaved output/state stores, hard-coded dimensions, and \texttt{\_\_launch\_bounds\_\_(32,1)}~\cite{kachua2026repo}. It also states that TVM FFI C binding was the largest speedup, that shared-memory staging hurt, and that the kernel was instruction-latency-limited rather than HBM-bandwidth-limited. UW SyFI followed a related decode shape: TVM FFI and destination passing, but with NVRTC plus the CUDA Driver API to avoid runtime-driver mismatch; the embedded decode kernel uses \texttt{V\_TILE=8}, one warp, float4 state movement, and warp-shuffle K reductions~\cite{syfi2026repo}. These two repositories suggest that top decode work treated the launch/runtime interface and one-warp small-kernel shape as first-class variables.

LLM-CUDA exposes the clearest explicit batch dispatch. Its retained decode entry point is \texttt{kernel.py::kernel\_hybrid\_dispatch}. The public summary routes \texttt{batch\_size <= 8} to a Triton recurrent path, with the B8 case using a one-warp, three-stage launch; \texttt{batch\_size == 16} goes to a vendored FlashInfer CuTe pretranspose path with a higher CTA count; \texttt{batch\_size == 48} goes to a dedicated vendored CuTe path with \texttt{TILE\_V=16} and four blocks per state; and the remaining large batches use the default vendored CuTe pretranspose path~\cite{llmcuda2026repo}. The same summary reports only small but stable improvements over its same-day baseline in several regimes, which is itself informative: decode was a regime-by-regime selection problem, not a single global kernel change.

This comparison reframes what \msinfer{} did not do. It did try split-factor changes, 256-thread large-batch variants, shared staging, CUDA graph replay, async prefetch, cluster q/k sharing, and cache-policy variants, and many were rolled back after full-workload regressions. However, it did not converge to a top-level dispatch table with separate small, B16, B48, and large-batch paths, nor to a pure one-warp TVM-FFI/Driver-API decode design like those in the top public repositories. Its NCU evidence of low issue-slot utilization, low achieved occupancy, and too few waves per SM is consistent with this explanation: B200 was underfilled, and the winning decode direction was not only ``do fewer arithmetic operations'' but ``choose the right launch/runtime/kernel shape for each batch regime.'' This explains the 1.50$\times$ decode gap to LLM-CUDA without overstating it as the main contest gap.

\subsection{Prefill Gap: Structural Versus Local Recurrence}

The prefill gap was not simply a smaller version of the decode gap. Decode is a tiny-latency regime where launch overhead, underfilled SMs, and baseline drift can dominate the interpretation. Prefill exposed a more algorithmic difference. Kachua, SyFI, and LLM-CUDA all show public evidence of chunk-level prefill paths~\cite{kachua2026repo,syfi2026repo,llmcuda2026repo}: Kachua uses chunkwise matmul plus Neumann/WY-style algebra and adaptive Split-WY or Flat-WY dispatch; SyFI uses a BF16 tensor-core WY-form chunked CUDA kernel with WMMA and \texttt{cp.async}; LLM-CUDA retained a Blackwell chunk kernel as its main path with a narrow short-shape fallback. These are not merely faster versions of \msinfer{}'s kernel. They change the unit of work from one token update at a time to chunk summaries and matrix-shaped subproblems.

\msinfer{}'s prefill log points in the opposite direction~\cite{jsg2026logs,bammuri2026agentprefill}. It improved a register-resident scalar recurrence through state tiling, loop fusion, V-dimension splitting, warp-parallel rows, algebraic fusion, launch-bounds tuning, gate pipelining, and split-factor dispatch. The same log also records a plateau: shared-memory q/k broadcast, fast math gates, higher split factors, shuffle reductions, packed stores, software prefetching, and q/k reduction decoupling either regressed or were flat. Its own research notes concluded that tensor cores, TMA, and TMEM were poorly matched to the current scalar recurrence, and that a step-change would require a chunked WY rewrite. This is an important nuance: within the scalar recurrence family, \msinfer{} had pushed hard; the missing piece was not one more cache hint, but leaving that family.

This explains the numerical asymmetry. Against LLM-CUDA's public retained artifacts, \msinfer{} was about 1.50$\times$ slower in decode but 4.61$\times$ slower in prefill. Against Kachua's public prefill trajectory rows near 61.17--61.50~$\mu$s, \msinfer{}'s 239.48~$\mu$s prefill was roughly 3.9$\times$ slower. We therefore treat decode as an important but secondary gap, and prefill as the main performance gap caused by missing structural reformulation, tensor-core exposure, and workload-specialized dispatch.

\section{Lessons Learned}

\subsection{Decode Was Shape-Dispatch; Prefill Was Structural}

The strongest lesson is that performance in the two tasks plateaued for different reasons. Decode improvements mattered, and the public repositories suggest that the missing decode move was stronger batch-regime dispatch and launch/runtime specialization. But public ratios show that \msinfer{} was much farther from top-three prefill than top-three decode. Future AI-assisted kernel systems should therefore ask two questions earlier: whether a microsecond-scale decode kernel needs separate shape-specific launch paths, and whether a prefill operator needs algorithmic reformulation before many iterations are spent on local recurrence micro-optimizations.

\subsection{Build Surface Is Part of Kernel Optimization}

Compiler target, language, binding, runtime/driver behavior, and destination-passing style can affect both performance and evaluator compatibility. SyFI's NVRTC/Driver API path and \msinfer{}'s \texttt{sm\_100a}/runtime-pin concerns point to the same conclusion: the evaluator-facing build surface is a real systems variable.

\subsection{Measurement Is an Optimization Surface}

The Modal/official divergence shows that benchmark configuration can steer the search toward or away from useful candidates. For microsecond-scale decode, repeated medians, isolated runners, per-workload views, FlashInfer/CUPTI-style pure GPU timing, and official-parity validation are not bookkeeping; they are part of the optimization method. In retrospect, \msinfer{}'s evaluation toolchain was useful but less centralized than the strongest public timing workflow: it reduced noise and added profiler context, but it did not fully replace the template path with an official-aligned kernel-latency harness.

\subsection{Algorithm and Measurement Co-Design}

The prefill gap and the measurement gap should not be treated as independent failures. Chunked and WY-style formulations introduce choices about sequence regime, chunk size, fallback path, and tensor-core-friendly tiling. Those choices are only useful if the benchmark loop can tell which shapes improved. The top public artifacts suggest that algorithm reformulation and evaluator-aligned timing co-evolved; \msinfer{} instead had strong local evidence collection but a weaker central signal for promoting workload-specific structural changes.

\subsection{Rejected Experiments Are Evidence}

The repo-local logs are useful precisely because they record failed hypotheses. They show that plausible local changes can be wrong for different reasons: decode cache-policy and cluster-sharing ideas could regress full-workload latency even when they looked reasonable on paper; decode NCU evidence showed that B200 was underfilled rather than purely bandwidth-saturated; prefill software prefetching added overhead despite high L1 hit rate; and q/k reduction decoupling could not overlap work that still executed in warp program order. Separate Full-Agent branches also showed a different failure mode: a candidate can be technically interesting yet still violate the evaluator's submission contract. These rejected experiments strengthen the case-study claim. The absence of stronger prefill performance was not simply a lack of trying small CUDA edits; the remaining gap was structural.

\subsection{Full-Agent-Style Companion Evidence}

Full-Agent experiments were also attempted, but an error in the final submission/evaluator path prevented an official result. We therefore use those artifacts only as evaluator-contract and search-process context, not as \agentassist{} performance evidence. The associated repo-local workflow skills---\texttt{research}, \texttt{optimize}, \texttt{bench}, and \texttt{log-result}~\cite{jsg2026workflow}---document search behavior and rejected hypotheses; official \agentassist{} performance evidence remains limited to decode \texttt{69f517b} and prefill \texttt{864f286}.

In the inspected \texttt{jsg1504/mlsys26} workspace, these skills made the optimization trace explicit. The loop read the current kernel, checked previous attempts, collected or requested profiler evidence, implemented one change, benchmarked it on Modal B200, and appended structured records. By the end of the inspected companion log, the prefill history contained 33 benchmark records, and the broader workspace also contained decode optimization records.

This loop made the search auditable, but its primary benchmark records were still Modal benchmark latency and speedup records. Later \msinfer{} agent repositories added more specialized analysis---NCU scorecards, surrogate scorecards, workload inventories, and direct phase timers---which improved diagnosis. The evidence base was therefore broad, but not organized around a single official-adjacent kernel-latency objective. That distinction helps explain why many local micro-optimizations were explored, while the larger workload-specific prefill reformulation path remained harder to evaluate and promote.

\begin{table}[t]
\caption{Repo-local companion automation evidence loop inspected for this case study. This is not official \agentassist{} submission evidence.}
\label{tab:evidence-loop}
\vskip 0.05in
\begin{center}
\begin{small}
\setlength{\tabcolsep}{3pt}
\begin{tabular}{>{\raggedright\arraybackslash}p{0.78in}>{\raggedright\arraybackslash}p{2.25in}}
\toprule
Unit & Evidence produced \\
\midrule
\texttt{research} & Reads kernel code, benchmark history, and optimization log; returns ranked ideas with expected impact and risk. \\
\texttt{optimize} & Runs a one-change loop: profile, select, implement, benchmark, profile again, then accept or revert. \\
\texttt{bench} & Packs the solution and runs Modal B200 benchmarks, recording status, latency, speedup, correctness, workload count, and notes. \\
\texttt{log-result} & Summarizes benchmark trends and optimization-history learnings from JSONL and Markdown logs. \\
\bottomrule
\end{tabular}
\end{small}
\end{center}
\vskip -0.1in
\end{table}

\subsection{AI Assistance Still Needs Systems Judgment}

Agents can generate code, summarize logs, and propose variants. They do not automatically know which measurements transfer to the official evaluator, whether a prefill recurrence needs a WY reformulation, or whether a package satisfies the evaluator's source-discovery contract. Human judgment remained central to deciding what evidence counted.

\section{Limitations}

This report is a case study, not a reproduced benchmark of all public submissions. We tie public comparison values to repository commits or official writeups. Several interpretations of root causes are based on logs and public comparison rather than isolated controlled experiments. We therefore use careful language: ``suggests,'' ``indicates,'' and ``is consistent with'' are often more accurate than causal absolutes.

This report intentionally keeps the main story focused on the \agentassist{} GDN submission. Full-Agent experiments belonged to a separate track and are not used as performance evidence for the \agentassist{} result. Where they are mentioned, we frame them only as cautionary examples about submission and evaluator contracts.

\section{Conclusion}

Our team, \msinfer{}, produced a correct \agentassist{} \gdn{} submission with an official 1.58$\times$ speedup in the MLSys 2026 FlashInfer Contest. Its value lies not in its final rank but in the boundary it exposes. AI-assisted micro-optimization, runtime pinning, and careful measurement can produce a useful kernel artifact. Yet the public top-three submissions show that for \gdn{} prefill, the top path required a larger algorithmic shift: chunked, shape-specialized, tensor-core-friendly recurrence computation. The broader lesson is that this algorithmic shift also required a reliable timing signal to select among workload-specific variants. In this sense, the kernel body was necessary but not sufficient.

\bibliographystyle{mlsys2026}
\bibliography{references}

\appendix

\section{Evidence Audit}

\begin{table*}[t]
\caption{Evidence sources and status for reported claims.}
\label{tab:claims-audit}
\vskip 0.05in
\begin{center}
\begin{scriptsize}
\setlength{\tabcolsep}{3pt}
\begin{tabular}{>{\raggedright\arraybackslash}p{2.55in}>{\raggedright\arraybackslash}p{1.0in}>{\raggedright\arraybackslash}p{2.65in}}
\toprule
Claim & Evidence class & Evidence status or source \\
\midrule
The final \agentassist{} \gdn{} submission by \msinfer{} achieved a 1.58$\times$ speedup & Official result & Final leaderboard or evaluator output available to the authors; reported here as official, not independently reproduced. \\
Contest GDN scoring averaged decode and prefill definition speedups, and speedup was measured against the simple definition reference rather than an optimized FlashInfer baseline & Contest material & Contest FAQ and starter-kit material~\cite{flashinfer2026starterkit}; the interpretation is scoped to that contest documentation. \\
Final decode latency 9.315~$\mu$s and prefill latency 239.48~$\mu$s & Official result & Per-definition final evaluator output available to the authors; reported as approximate averages. \\
Decode passed 54/54 and prefill passed 100/100 workloads & Official / artifact & Final validation logs for the submitted decode and prefill artifacts. \\
LLM-CUDA reported 6.201~$\mu$s decode and 51.992~$\mu$s prefill & Public report & LLM-CUDA writeup and retained-artifact repository~\cite{llmcuda2026agentassist,llmcuda2026repo}; not reproduced here. \\
Kachua public repository records selected decode variants at 5.44--5.47~$\mu$s median-of-workload medians and prefill rows at 61.17--61.50~$\mu$s & Public repo & Kachua decode README and prefill \texttt{results.csv}~\cite{kachua2026repo}; treated as public trajectory numbers, not official leaderboard averages. \\
Kachua decode source and README describe direct TVM FFI C binding, one-warp \texttt{BV=8} CUDA, vectorized loads, paired butterfly reductions, and \texttt{\_\_launch\_bounds\_\_(32,1)} & Public repo & Kachua decode README and \texttt{kernel.cu}~\cite{kachua2026repo}. \\
Kachua prefill source uses chunkwise matmul, Neumann/WY-style algebra, adaptive \texttt{CHUNK}/\texttt{BV}, Split-WY/Flat-WY dispatch, and tensor-core \texttt{tl.dot} contractions & Public repo & Kachua prefill \texttt{kernel.py}, \texttt{results.csv}, and experiment log~\cite{kachua2026repo}. \\
SyFI used NVRTC/Driver API, TVM FFI, DPS, a one-warp \texttt{V\_TILE=8} decode kernel, and a BF16 tensor-core WY-form chunked prefill kernel with WMMA/\texttt{cp.async} & Public repo/report & SyFI source, configs, and writeup~\cite{syfi2026agentassist,syfi2026repo}; no raw latency table is reported here. \\
LLM-CUDA retained decode used hybrid dispatch: \texttt{B <= 8} recurrent path, B16 high-CTA CuTe path, B48 \texttt{TILE\_V=16} path, and default large-batch CuTe path & Public repo/report & LLM-CUDA retained decode summary and source~\cite{llmcuda2026agentassist,llmcuda2026repo}. \\
LLM-CUDA retained prefill replaced wide pair dispatch with a Blackwell chunk kernel as the main path and a narrow short-shape fallback & Public repo/report & LLM-CUDA retained prefill summary and source~\cite{llmcuda2026agentassist,llmcuda2026repo}. \\
Modal vs official decode divergence values & Internal / official & Internal Modal logs and official extra-round evidence available to the authors; used as an evaluator-alignment example, not a controlled causal experiment. \\
Kachua repository contains \texttt{bench\_fi\_timing.py}, \texttt{bench\_fi\_timing\_modal.py}, and program notes that distinguish \texttt{kernel\_latency} from \texttt{benchmark\_latency} and prefer the former & Public repo & Kachua timing scripts and repository notes~\cite{kachua2026repo}. \\
\msinfer{} agent repositories and companion JSG workflow tooling logged Modal benchmark latency/speedup histories and later added NCU, surrogate-scorecard, workload-inventory, and phase-timer analysis; this audit did not find a single FlashInfer/CUPTI-style kernel-latency objective used as the central optimization primitive & Artifact / interpretation & Inspected \texttt{jsg1504/mlsys26} skills/logs, \msinfer{} agent scripts, benchmark logs, and profiler artifacts; negative claim is limited to the inspected artifacts. \\
Algorithm reformulation and evaluator-aligned timing appear to have reinforced each other in public high-performing prefill workflows & Interpretation & Derived from public repository comparison and timing-script evidence; not a controlled ablation or same-harness reproduction. \\
Repo-local companion automation loop used \texttt{research}, \texttt{optimize}, \texttt{bench}, and \texttt{log-result} workflow units; this loop is not official \agentassist{} submission evidence & Artifact / workflow context & \texttt{jsg1504/mlsys26} workflow skills and logs~\cite{jsg2026workflow,jsg2026logs}; companion context only. \\
Decode artifact is the \agentassist{} \texttt{gdn-decode} state at \texttt{69f517b} & Artifact & MSInfer Agent-Assisted decode repository metadata~\cite{bammuri2026agentdecode}. \\
Decode config uses \texttt{language = "cuda"}, \texttt{decode\_submit\_entry.py::run}, source files \texttt{kernel.cu} and \texttt{decode\_submit\_entry.py}, and destination-passing style & Artifact & Checked against the local Agent-Assisted decode \texttt{config.toml} and source tree. \\
\msinfer{} accepted decode selected \texttt{split\_factor} by \texttt{batch\_size} inside one templated CUDA kernel family rather than a public high-performing hybrid dispatch table & Artifact / interpretation & Agent-Assisted decode \texttt{kernel.cu}, rejected-experiment logs, and public decode-source comparison; not a claim of exhaustive search. \\
Prefill artifact is the \agentassist{} \texttt{gdn-prefill} state at \texttt{864f286} & Artifact & MSInfer Agent-Assisted prefill repository metadata~\cite{bammuri2026agentprefill}. \\
Prefill config uses \texttt{language = "python"}, \texttt{msinfer\_entry.py::run}, and destination-passing style for \texttt{gdn\_prefill\_qk4\_v8\_d128\_k\_last} & Artifact & Checked against the local Agent-Assisted prefill \texttt{config.toml} and source tree. \\
Full-Agent experiments were attempted, but an error in the final submission/evaluator path prevented an official result; those artifacts are excluded from \agentassist{} performance evidence & Internal / artifact & Internal validation or submission records; included only as evaluator-contract context, not \agentassist{} performance evidence. \\
Companion prefill benchmark history contained 33 records & Artifact / workflow context & Verified from \texttt{logs/prefill/bench\_history.jsonl}; companion evidence, not official \agentassist{} submission evidence. \\
Companion prefill logs show early large local-recurrence gains followed by reverted or flat micro-optimizations and an explicit scalar-recurrence plateau & Internal measurement / interpretation & Prefill optimization log and benchmark-history entries~\cite{jsg2026logs}; companion evidence only. \\
Prefill reformulation was the largest visible gap & Interpretation & Supported by public ratios and design comparison; not a same-harness reproduction of all submissions. \\
\bottomrule
\end{tabular}
\end{scriptsize}
\end{center}
\vskip -0.1in
\end{table*}

\end{document}